\documentclass[11pt]{article}
\usepackage{moriond,epsfig}

\def\beq{\begin{equation}}
\def\eeq{\end{equation}}
\def\bea{\begin{eqnarray}}
\def\eea{\end{eqnarray}}
\def\bq{\begin{quote}}
\def\eq{\end{quote}}

\def\gappeq{\mathrel{\rlap {\raise.5ex\hbox{$>$}}
{\lower.5ex\hbox{$\sim$}}}}

\def\lappeq{\mathrel{\rlap{\raise.5ex\hbox{$<$}}
{\lower.5ex\hbox{$\sim$}}}}

\def\be{\begin{equation}}
\def\ee{\end{equation}}
\def\bea{\begin{eqnarray}}
\def\eea{\end{eqnarray}}

\begin{document}
\vspace*{4cm}
\title{CONSTRAINTS ON LEPTON FLAVOUR VIOLATION AND LEPTOGENESIS}

\author{ M. RAIDAL }

\address{ National Institute of Chemical Physics and Biophysics,
Tallinn 10143, Estonia}

\maketitle\abstracts{
We consider the scenario of direct leptogenesis from the decays of
heavy sneutrino which also plays a role of inflaton. 
We calculate its predictions for flavour-violating
decays of charged leptons. We find that $\mu \to e \gamma$ should occur
close to the present experimental upper limits, as might also $\tau \to
\mu \gamma$.
}

\vspace*{1cm}
\noindent


\section{Introduction}

Leptogenesis \cite{fy} is currently the favorite scenario for explaining 
the observed baryon asymmetry  of the Universe. It is related to the light
neutrino masses and mixings via the seesaw mechanism~\cite{seesaw}. 
In supersymmetric models all the neutrino observables can also be related to 
the lepton flavour violating (LFV) decays of charged leptons which are induced
by the heavy singlet neutrino Yukawa couplings via the renormalization effects. 
Implications of light neutrino masses on the heavy neutrino masses and
couplings~\cite{thomas}, and the relations between thermal leptogenesis 
and LFV~\cite{sasha} have been extensively studied. The appeal and popularity
of the thermal leptogenesis comes form its predictive power.
On the other hand, the second completely predictive leptogenesis scenario -
direct leptogenesis form the scalar field which dominates the 
Universe~\cite{sn2,us2,us} -
has got somewhat less attention. 

The purpose of this talk is to review the most predictive scenario of
direct leptogenesis in which the lightest heavy singlet sneutrino
plays the role of inflaton~\cite{sn1}.     
Inflation~\cite{inf} has become the paradigm for early cosmology, particularly
following the recent spectacular CMB data from the WMAP satellite~\cite{wmap}, which
strengthen the case made for inflation by earlier data, by measuring an
almost scale-free spectrum of Gaussian adiabatic density fluctuations
exhibiting power and polarization on super-horizon scales, just as
predicted by simple field-theoretical models of inflation.
Ever since inflation was proposed, it has been a puzzle how to integrate 
it with ideas in particle physics. For example, a naive GUT Higgs field 
would give excessive density perturbations, and no convincing concrete 
string-theoretical model has yet emerged. 
In this conceptual vacuum, 
models based on simple singlet scalar fields have held sway~\cite{inf}. 
Here we assume the simplest scenario 
in which the lightest heavy singlet sneutrino drives inflation. This 
scenario constrains in interesting ways many of the 18 parameters of 
the minimal seesaw model for generating three non-zero light neutrino 
masses.

This minimal sneutrino inflationary scenario (i) yields a simple ${1 \over
2} m^2 \phi^2$ potential with no quartic terms, with (ii) masses $m$ lying
naturally in the inflationary ballpark. The resulting (iii) spectral index
$n_s$, (iv) the running of $n_s$ and (v)  the relative tensor strength $r$
are all compatible with the data from WMAP and other 
experiments~\cite{wmap}.
Moreover, fixing $m \sim 2 \times 10^{13}$~GeV as required by the observed
density perturbations (vi) is compatible with a low reheating temperature
of the Universe that evades the gravitino problem~\cite{gr}, (vii) realizes
leptogenesis~\cite{us2,us} in a calculable and viable way, (viii) constrains neutrino
model parameters, and (ix) makes testable predictions for the
flavour-violating decays of charged leptons~\cite{us}.

\section{Reheating and Leptogenesis in our Scenario}

We assume the chaotic inflation with a $V = {1 
\over 2} m^2 
\phi^2$ potential - the form expected for a heavy singlet sneutrino - in 
light of WMAP~\cite{wmap}. Defining $M_P \equiv 1/\sqrt{8 \pi G_N} 
\simeq 2.4 \times 10^{18}$~GeV, the conventional slow-roll inflationary 
parameters are
\beq
\epsilon \equiv {1 \over 2} M_P^2 \left( {V^\prime \over V} \right)^2 = 
{2 M_P^2 \over \phi_I^2}, \;
\eta \equiv M_P^2 \left( {V^{\prime\prime} \over V} \right) = {2 M_P^2 
\over \phi_I^2}, \;
\xi \equiv M_P^4 \left( {V V^{\prime\prime\prime} \over V^2} \right) = 
0,
\label{slowroll}
\eeq
where $\phi_I$ denotes the {\it a priori} unknown inflaton field value 
during inflation at a typical CMB scale $k$. 
Assuming $N=50$ e-foldings of inflation,  the  WMAP data implies the
result~\cite{us}
\beq
m \; \simeq 1.8 \times 10^{13}~{\rm GeV}.
\label{phimass}
\eeq
This is comfortably within the range of heavy 
singlet (s)neutrino masses usually considered, namely $m_N \sim 10^{10}$ 
to $10^{15}$~GeV.

Is this simple $\phi^2$ sneutrino model compatible with the WMAP data? The 
primary CMB observables are the spectral index
\beq
n_s = 1 - 6 \epsilon + 2 \eta = 1 - {8 M_P^2 \over \phi^2_I} \simeq 
0.96,
\label{ns}
\eeq
the tensor-to scalar ratio
\beq
r \equiv {A_T \over A_S} = 16 \epsilon = {32 M_P^2 \over \phi^2_I} \simeq 
0.16,
\label{r}
\eeq
and the spectral-index running
\beq
{d n_s \over d {\rm ln} k} = {2 \over 3} \left[ \left( n_s - 1 \right)^2 - 
4 \eta^2 \right] + 2 \xi =   {32 M_P^4 \over \phi^4_I}  \simeq  8 \times 10^{-4}.
\label{values}
\eeq
The value of $n_s$ extracted from WMAP data depends whether, for example,
one combines them with other CMB and/or large-scale structure data. 
At the moment this scenario appears to be compatible with all data.

Following the  inflationary epoch, reheating of the Universe results from
the decays of inflaton sneutrino. 
Assuming, as usual, that the sneutrino inflaton decays 
when the the Hubble expansion rate $H \sim m$, and that the expansion rate 
of the Universe is then dominated effectively by non-relativistic matter 
until $H \sim \Gamma_\phi$,  where $\Gamma_\phi$ is the inflaton decay width, 
we estimate
\beq
T_{RH} = \left( 
\frac{90}{ \pi^2 g_*} 
\right)^{1 \over 4} \sqrt{\Gamma_\phi M_P} ,
\label{TRH}
\eeq
where $g_*$ is the number of effective relativistic degrees of freedom in 
the reheated Universe.
In the minimal sneutrino inflation scenario considered here we have
$\phi\equiv \tilde N_1,$ $m \equiv M_{N_1}$ and 
\bea
\Gamma_\phi\equiv \Gamma_{N_1} = 
\frac{1}{4 \pi} (Y_\nu Y_\nu^\dagger)_{11} M_{N_1},
\eea
where $Y_\nu$ is the neutrino Dirac Yukawa matrix.
If the relevant neutrino Yukawa coupling 
$(Y_\nu Y_\nu^\dagger)_{11} \sim 1$, the previous choice $m=M_{N_1} \simeq 
2 \times 10^{13}$~GeV would yield $T_{RH} > 10^{14}$~GeV, considerably greater 
than $m$ itself. Such a large value of
$T_{RH}$ would be {\it very problematic} for the thermal production of
gravitinos~\cite{gr}, and leads to thermal leptogenesis (for the alternative
leptogenesis solution to the gravitino problem see~\cite{soft}). 
However, in the light of present favourite range
of light neutrino masses one can naturally guess from the seesaw
mechanism 
$(Y_\nu Y_\nu^\dagger)_{11}\ll 1$. In this case $T_{RH}$ could be much lower,
usually below the lightest sneutrino mass $2\times 10^{13}$ GeV,
leading to direct leptogenesis.

To calculate the lepton asymmetry to entropy density
ratio $Y_L=n_L/s$ in inflaton decays we need to know the
produced entropy density
\bea
s=\frac{2\pi^2}{45} g_* T_{RH}^3,
\eea
and to take into account that inflaton dominates the
Universe. In this case one obtains~\cite{sn2}
\bea
Y_L=\frac{3}{4} \epsilon_1 \frac{T_{RH}}{M_{N_1}},
\label{yl}
\eea
where $\epsilon_1$ is the CP asymmetry in 
$\phi\equiv \tilde N_1$ decays.
We consider now the most constrained scenario in which the inflaton is the
lightest sneutrino, which requires $M_{N_3} > M_{N_2} > M_{N_1} \simeq 2 \times
10^{13}$~GeV. This implies that our problem is completely characterized by 
 only one parameter, either $\tilde m_1$ or $T_{RH}$.
The observed baryon asymmetry of the
Universe gives a lower bound on
the reheating temperature $T_{RH}>10^6$ GeV.

\section{Leptogenesis Predictions for Lepton Flavour Violation}

In this Section, we make predictions on the
LFV decays  $\mu \to e \gamma$ and $\tau \to \mu \gamma$ in this scenario. 
We first calculate neutrino Yukawa
couplings using the parametrization in terms of the light and heavy
neutrino masses, mixings and the orthogonal parameter matrix given
in~\cite{ci}. This allows us to calculate exactly the baryon asymmetry of
the Universe from the CP asymmetry $\epsilon_1$~\cite{vissani} and the
reheating temperature of the Universe $T_{RH}$ via eq.(\ref{yl}). 
For neutrino parameters
yielding successful leptogenesis, we calculate the branching ratios of LFV
decays~\cite{lfv}. For all calculational details we refer the reader 
to~\cite{us,cp,er}.

\begin{figure}[t]
\centerline{\epsfxsize = 0.5\textwidth \epsffile{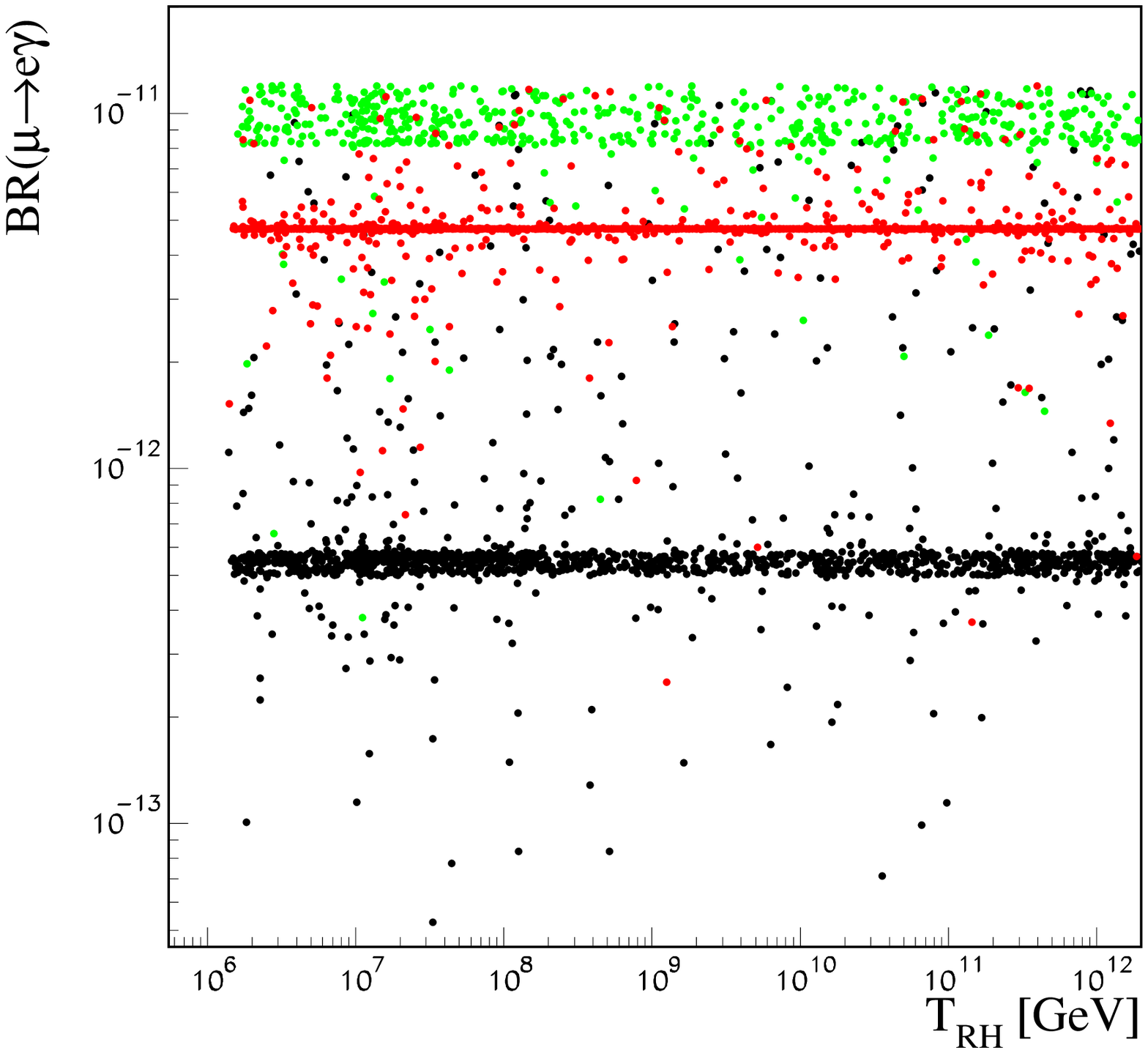}
\hfill \epsfxsize = 0.5\textwidth \epsffile{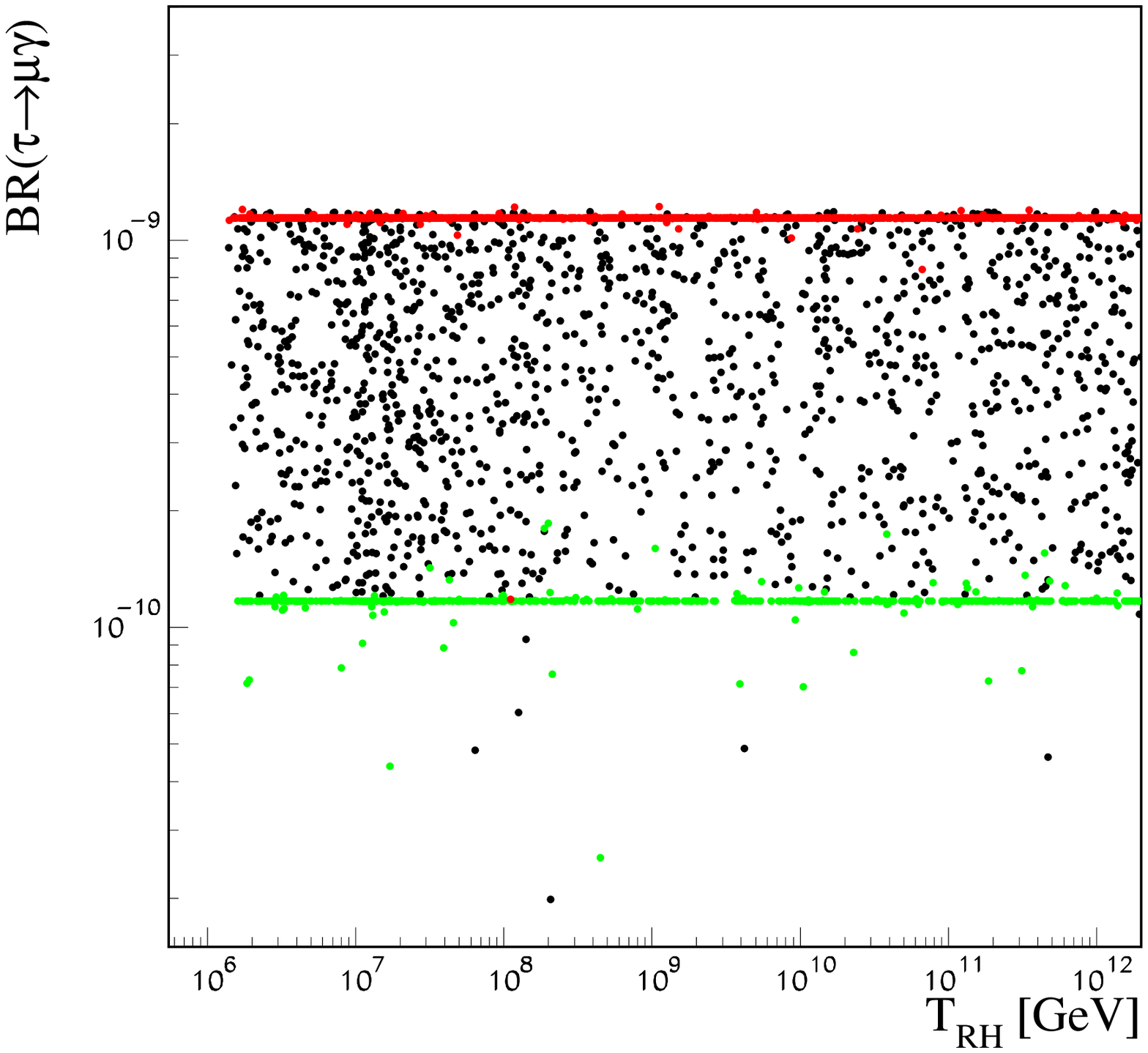}
}
\caption{\it
Calculations of BR$(\mu \to e \gamma)$ and BR$(\tau \to \mu \gamma)$
on left and right panels, respectively. Black points correspond to
$\sin \theta_{13} = 0.0$,  $M_2 = 10^{14}$~GeV, 
and $5 \times 10^{14}$~GeV $< M_3 < 5 \times 10^{15}$~GeV.
Red points correspond to $\sin \theta_{13} = 0.0$,  $M_2 = 5 \times 10^{14}$~GeV, 
and  $M_3 = 5 \times 10^{15}$~GeV, while green points correspond to
 $\sin \theta_{13} = 0.1$, 
$M_2 = 10^{14}$~GeV, and $M_3  = 5 \times 10^{14}$~GeV.
\vspace*{0.5cm}}
\label{fig3}
\end{figure}

The results on the branching ratios are presented in  Fig.~\ref{fig3}
where we plot BR$(\mu \to e \gamma)$ (panel (a)) and  BR$(\tau \to \mu \gamma)$
(panel (b)) against the reheating temperature of the Universe $T_{RH}.$
We see immediately that values of $T_{RH}$
anywhere between $2 \times 10^6$~GeV and $ 10^{12}$~GeV are
attainable in principle. The lower bound is due to the lower
bound on the CP asymmetry~\cite{di2}, while the upper bound comes
from the gravitino problem. 
The black points in panel (a)  correspond to the choice 
$\sin \theta_{13} = 0.0$,  $M_2 = 10^{14}$~GeV, 
and $5 \times 10^{14}$~GeV $< M_3 < 5 \times 10^{15}$~GeV.
The red points correspond to $\sin \theta_{13} = 0.0$,  $M_2 = 5 \times 10^{14}$~GeV, 
and  $M_3 = 5 \times 10^{15}$~GeV, while the green points correspond to
$\sin \theta_{13} = 0.1$, $M_2 = 10^{14}$~GeV, and $M_3  = 5 \times 10^{14}$~GeV.
The soft SUSY breaking parameters are fixed as 
$(m_{1/2}, m_0) = (800,170)$~GeV
which is the upper bound for providing the cold dark matter of the 
Universe~\cite{dm}, and  and $\tan\beta=10$.
We see a very striking narrow, densely populated  bands for BR$(\mu \to e \gamma)$, 
with some outlying points at both larger and smaller
values of BR$(\mu \to e \gamma)$. 
The width of the black band is due to variation of $M_{N_3}$ showing that  
BR$(\mu \to e \gamma)$ is not very sensitive to it. However, 
BR$(\mu \to e \gamma)$ strongly depends on $M_{N_2}$ and  $\sin \theta_{13}$
as seen by the red and green points, respectively. Since
BR$(\mu \to e \gamma)$ scales approximately as $m_{1/2}^{-4}$, the lower
strip for $\sin \theta_{13} = 0$ would move up close to the experimental
limit if $m_{1/2} \sim 500$~GeV, and the upper strip for $\sin \theta_{13}
= 0.1$ would be excluded by experiment.

Panel (b) of Fig.~\ref{fig3}
 shows that BR$(\tau \to \mu \gamma)$ depends strongly on 
 $M_{N_3}$, while the dependence on $\sin \theta_{13}$ 
and  on $M_{N_2}$ is negligible. 
The numerical values of BR$(\tau \to \mu \gamma)$
 are somewhat below the present experimental upper
limit BR$(\tau \to \mu \gamma) \sim 10^{-7}$, but we note that the results
would all be increased by an order of magnitude if $m_{1/2} \sim 500$~GeV.
In this case, panel (a) of Fig.~\ref{fig3} tells us that the experimental
bound on BR$(\mu \to e \gamma)$ would enforce $\sin \theta_{13} \ll 0.1$,
but this would still be compatible with BR$(\tau \to \mu \gamma) >
10^{-8}$.

As a result, Fig.~\ref{fig3} strongly suggests that fixing the observed
baryon asymmetry of the Universe for the direct sneutrino leptogenesis
($T_{RH}<2\times 10^{12}$~GeV $<M_{N_1}$) implies a prediction for the LFV
decays provided $M_{N_2}$ and/or $M_{N_3}$ are also fixed.

\section{Conclusions}

The main results of our scenario are the following.
{\it First}, reheating of the Universe is
due to the neutrino Yukawa couplings, and therefore can be
related to light neutrino masses and mixings. 
{\it Secondly},  
the lepton asymmetry is created in direct 
sneutrino-inflaton decays.
There is only one parameter describing the efficiency of
leptogenesis in this minimal sneutrino inflationary scenario in all
leptogenesis regimes - the reheating temperature of the Universe - to 
which
the other relevant parameters can be related.
This should be compared with the general thermal 
leptogenesis case
which has two additional independent parameters, 
namely the lightest heavy neutrino mass and width. 
{\it Thirdly},
imposing the requirement of successful leptogenesis and correct amount of
cold dark matter, we can predict
branching ratios for $\mu\to e\gamma$ and $\tau\to\mu\gamma$
in a very narrow band within
about one order of magnitude of the present experimental bound.

\section*{Acknowledgments}

We thank J. Ellis and T. Yanagida
for collaboration and discussions.
This work is partially supported  by the ESF grant No. 5135.

\end{document}